\documentclass[11pt]{article}

\newcommand{\lbl}[1]{\label{#1}}
\newcommand{\bib}[1]{\bibitem{#1}}

\usepackage{amsxtra}
\usepackage{amsmath}
\usepackage{graphics}

\usepackage{graphicx}
\graphicspath{%
    {converted_graphics/}
    {/}
}
\begin{document}


\title{Radial distance on a stationary frame in a homogeneous and isotropic universe.}
\author{Robert C. Fletcher\footnote{Mailing address: 1000 Oak Hills Way, Salt Lake City, UT 84108.  robert.c.fletcher@utah.edu, robcfletcher@comcast.net}
\\
{\small Bell Telephone Laboratories (ret)}\\
{\small Murray Hill, New Jersey.}}  
\date{}

\maketitle

\begin{quote} {\small \begin{center} {\bf Abstract}\end{center}

This paper presents a physical distance to all radial events in a homogeneous and isotropic universe as a transform from Friedman-Lemaitre-Robertson-Walker (FLRW) coordinates, the model that solves
 the Einstein Field equation for an ideal fluid.  
Any well behaved transform is also a solution.  The problem is relating the coordinates of the transform to observables.  In the present case the objective is to find $T,R$ on a stationary frame that 
has the $R$ be a physical observable for all distances.  We do this by working backwards, assuming the form of the metric that we desire, with some undetermined coefficients.  These coefficients are then related to the partial derivatives of the transform.  The transformed coordinates $T,R$ are found by the integration of partial differential equations in the FLRW variables.  We show that $dR$ represents a reading on the same type of physical ruler as the radial differential of the FLRW metric,  
which makes it observable. We develop a criterion for how close the transformed $T$ comes to an observable time. Close to the space origin at the present time, $T$ also becomes physical, so that the stationary acceleration becomes Newtonian. We show that a galactic point on a $R,T$ plot starts close to the space origin at the beginning, moves out to a physical distance and finite time where it can release light that will be seen at the origin at the present time.  Lastly, because the observable $R$ has a finite limit at a finite $T$ for $t = 0$ where the galactic velocity approaches the light speed, we see that the universe filled with an ideal fluid as seen on clocks and rulers on the stationary frame has a finite extent like that of an expanding empty universe, beyond which are no galaxies and no space. \\\

}\end{quote}

keywords: cosmology theory---distances and redshifts\\\

\tableofcontents

\section{Introduction}

The FLRW metric\cite{R}\cite{Wa} metric is derived for a homogeneous and isotropic universe.   With the assumption of a stress-energy tensor for an ideal liquid, the solution of the Einstein Field Equation
gives the FLRW model of the universe\cite{P}\cite{MTW}\cite{We1} used in this paper.  The FLRW variables ($t,\chi,\theta,\phi$) are interpreted as commoving coordinates with $dt$ being a physical (observable) time, and $a(t)d\chi$ being a physical radial distance, and $a(t)$ is the scale factor of the FLRW universe.
We visualize these galactic points representing galaxies as moving away from an observer (us) at the origin at a velocity measured by distance and time on our stationary frame. We can measure their velocity by the red shift of the lines of their spectrum, but their distance is more difficult.  The proper distance $a(t)\chi$, the distance to $\chi$ at constant $t$, does not have a Minkowski metric with $t$ (except for small $\chi$) required for physical coordinates.
Weinberg describes a sequence of physical distance measures, running from parallax measurements for nearby stars to standard candle measurements for more distant sources\cite{We1}.  These all accurately represent the observables as derived from
 the FLRW coordinates, but are not our stationary distance; they differ from each other and from proper distance for large red shifts.

This paper presents a physical distance on a stationary frame to all radial events in a homogeneous and isotropic universe as a transform from FLRW coordinates.  Any well behaved transform from FLRW coordinates is also a solution of the field equation because tensor equations are invariant to transforms.  The problem is relating the coordinates of the transform to observables. In the present case we would like the observables to be the time on clocks and the distance on rulers on the stationary frame attached to the spatial origin,  but this is only possible close to the origin.  The next best objective is to find $T,R$ for all distances that become these observables close to the origin, and then have the $R$ be a physical observable for all distances.  

The key to showing that the distance is universal is to show that it is rigid with the same constant units as the FLRW radial differential distance on galactic points, which units are assumed to be physical.   
We outline here how we will do this.  We will look for a transform to radial coordinates $T,R$ from radial FLRW coordinates $t,\chi$ that are attached to the FLRW origin $\chi = 0$.  
We can call $T,R$ stationary because the velocity of the galactic point is zero at $\chi = 0$. 
 
To find the transform of $t,\chi$ to $T,R$ we will start with a metric with $dT,dR$ that is spherically symmetric in space, and find the coefficients of $dT,dR$ by putting constraints on them and then integrating the resultant $dT,dR$, subject to the boundary conditions at $\chi = 0$.  The constraints we will impose is that the points of $R$ be motionless with respect to each other (i.e., rigid), that the the line element $ds$ of $dT,dR$ be the same as FLRW, that its metric (the line element expressed in coordinate differentials) have no cross products $dTdR$ (i.e., diagonal), that the metric in $dT,dR$ become Minkowski close to the origin, and that $dR$ have the same units for all distances as the radial differential of the FLRW line element, $a(t)d\chi$.
The resulting paths of galactic points and photons in the stationary coordinates closely resembles the known transform to $T,R$ from $t,\chi$ for an expanding empty universe\cite{R2}.  These latter have a Minkowski metric for all distances, and have universal physical coordinates $T,R$.

\section{Assumptions}

We will use the word "line element" to represent the invariant $ds$ and the word "metric" for the $ds$ of a particular set of coordinates, and limit ourselves to the time and radial distance of spherical space symmetry (no transverse motion). The analysis depends on the following assumptions: \\

 (i) The FLRW metric is a valid representation of a homogeneous and isotropic universe, whose coordinates ($\chi, \theta,\phi.t$) are co-moving with the galactic points (representing smoothed out galaxies), and whose  universe scale factor is $a(t)$. \\

Definitions: A ``rigid frame'' is a plane where at any given time none of its spatial points has movement compared to any other spatial point. 
We will call ``physical'' those coordinate whose differentials over some interval of time or distance that are the same at any two inertial points along a rigid frame connecting them.
 Moving rigid frames will generally have different physical coordinates from each other, although all will use the same invariant units when representing clock and ruler readings.  We call these coordinates physical because they represent those physical clocks and rulers on an inertial rigid frame on earth. For instance, the physical clock might be the spectral frequency of a  standard atom and the physical ruler be its wavelength.  Their physical differentials will be related by an invariant Minkowski metric $\hat{M}$.  Any other coordinate that represents the reading on one of these physical instruments we will also call physical  (see assumption iii).
  Physical velocity of a moving point is the ratio of the differential physical distance that the point moves in a differential physical time when both time and distance coordinates are at the same location as the moving point. 
 Our objective is to find a $\ physical$ distance for all radial events in the universe.  \\

ii) ``The radial coordinate differentials $dt$ and $a(t)d\chi$ of FLRW are assumed to be differentials of physical coordinates.'' \\

 This is a usual assumption made plausible because they have the Minkowski metric in these dimensions so that the cosmic point $\chi$ is assumed to be on a rigid inertial frame.  
$dt$ will represent the time increment on a galactic clock that keeps time like a standard clock, and $a(t)d\chi$ will represent the physical distance between adjacent galactic points separated by $d\chi$, as measured on a  local standard ruler comoving with these galactic points.
  \\

Definition:  The coordinates $x^\mu$ ($R,\theta,\phi,T$) are transforms from FLRW with spherically symmetric space coordinates
whose space origins are the same as FLRW, so $\chi = 0$ and $V = 0$ at $R = 0$.  We will therefore call them stationary coordinates.  Like all well behaved transforms, they satisfy the Einstein field equations, but we need to relate them to observables.  The physical velocity with respect to a galactic point $t,\chi$ of a point on $R$ will be 
$V = a(t)(\partial \chi/\partial t)_R$.   
  For this radial point $R$ of $x^\mu$, we can find contravariant velocity vectors  $U^\mu = dx^\mu /ds$ and acceleration vectors $A^\mu = DU^\mu /Ds$ whose components will transform the same as $dx^\mu$.  $R$ is rigid because the radial component of $U^\mu$ in stationary coordinates is $(\partial R/\partial T)_R(dT/ds) \equiv 0$, so the points of $R$ are motionless with respect to each other. \\

($iii$) ``We assume an extension of the Bernal criteria\cite{B} that one of two observers will have a physical coordinate when the other does if each calculates the other's differential of that coordinate at the same space-time point compared to his own and finds these cross measurements to be equal.''\\  

This seems reasonable since this is the characteristic of the transform between moving inertial frames.
We will find a radial AP transform $(T,R)$ 
whose differential dR is physical for all distances by virtue of this assumption (Section \ref{V} and Appendix \ref{pd}). If the $dR$ represent readings on a stationary standard ruler, all on the same frame, they can be integrated to $R$ to determine a physical distance measured on a rigid physical ruler out into the far reaches of the universe.  \\



(note: Even though $a(t)d\chi$ is physical, when it is integrated at constant $t$, the  resultant integral $s_p = a(t)\chi$, usually called ``proper distance'' to the galactic point $\chi $, is not physical because the $\chi$'s are on different planes moving with respect to each other; i.e., $s_p$ is not rigid.  
That it is clearly not physical can also be seen from the ``proper'' velocity  of a galactic point $ v_p = (\frac {\partial s_p}{\partial t})_\chi = \chi \frac {da(t)}{dt} $, which increases linearly with distance from the origin, passing $c$ on its way to infinity for far distant galactic points, violating special relativity.
) \\

iv)  ``The known solution of the field equations of general relativity for the FLRW universe gives $a(t)$.''\\

Because of the properties of Riemann tensors, any transform from FLRW will also be a solution.  This allows evaluation of $T,R$ of an event for a given universe energy density.

\section{Procedure for finding stationary coordinates using the velocity $V$} 

We assume that the concentrated lumps of matter, like stars and galaxies, can be averaged to the extent that the universe matter can be considered continuous, and that the surroundings of every point in space can be assumed isotropic and the same for every point.

By embedding a maximally symmetric (i.e., isotropic and homogeneous) three dimensional sphere, with space dimensions $r$, $\theta$, and $\phi$, in a four dimension space which includes time $t$, one can obtain a differential line element $ds$ \cite[page 403]{We1} such that
\begin{equation}
	ds^2 = c^2dt^2 - a(t)^2 \left[ \frac {dr^2}{1-kr^2} +r^2 d\theta^2 +r^2 sin^2\theta d \phi ^2 \right],
\end{equation}
where 
\begin{equation}
	r = \left \{ \begin{array}{l} \sin{\chi},~~~k = 1,\\ \chi,~~~~k = 0,\\ \sinh{\chi},~~~k = -1,
	\end{array} \right.
\lbl{2"} 
\end{equation}
$k$ is a spatial curvature determinant to indicate a closed, flat, or open universe, resp., and 
\begin{equation}
d\chi^2 \equiv dr^2/(1-kr^2).
\end{equation}
$a(t)$ is the cosmic scale factor multiplying the three dimensional spatial sphere, and the differential radial distance is $a(t)d\chi$. 
 
The resulting equation for the differential line element becomes the FLRW metric: 
\begin{equation}
	ds^2 = c^2dt^2 - a(t)^2[d\chi^2 + r^2d\omega^2], 
\lbl{1}
\end{equation} 
where $d\omega^2 \equiv d\theta^2 + sin^{2} \theta d\phi^2 $.   
For radial world lines this metric becomes  Minkowski in form
with a differential of physical radius of $a(t)d\chi$.

We would now like to find radial transforms that will hold for all values of the FLRW coordinates and have a Minkowski metric for small distances from the origin as required if $dT$ and $dR$ are to be the time and distance on our stationary frame. 
The most general line element for a time dependent spatially spherically symmetric (i.e., isotropic) line element\cite{We1}  is
\begin{equation}
	ds^2 = c^2 A^2 dT^2 - B^2 dR^2  - 2cCdTdR- F^2(d\theta^2 + sin^2 \theta d\phi^2)
\lbl{5'}
\end{equation}
where $A$, $B$, $C$, and $F$ are implicit function of $T$ and $R$, but explicit functions of $t$ and $\chi$.  

We would like $T$ and $R$ to locate the same radial events as $t$ and $\chi$, so $T = T(t,\chi)$ and $R = R(t,\chi)$. 
therefore 
\begin{equation} \begin{array}{l}
dT = T_t dt + T_\chi d\chi = \frac{1}{c}T_t d\hat{t} + T_\chi d\chi , \\[2mm]
dR = R_t dt + R_\chi  d\chi = \frac{1}{c}R_t d\hat{t} + R_\chi d\chi,
\end{array}  \lbl{7} \end{equation}
where the subscripts  indicate partial derivatives
with respect to  the subscript variable and $d\hat{t} = cdt$.

We will look for transformed coordinates which have their origins on the same galactic point as $\chi =0$, so $R = 0$ when $\chi = 0$, where there will be no motion between them, and where $T $ is $t$, since the time on clocks attached to every galactic point
is $t$, including the origin.  We will keep the same angular coordinates for the transform and make $F = ar$ to correspond to the FLRW metric, but will find only radial transforms where the angular differentials are zero.  

Let us consider a radial point at $R$.  When measured from the FLRW system, it will be moving at a velocity given by
\begin{equation}
 V = a(t)\left( \frac{\partial \chi}{\partial t} \right)_{\! R} \equiv c\hat{V}, 
\lbl{12}
\end{equation}
This velocity will be the key variable that will enable us to obtain radial transforms of the full radial coordinates.  

We will first use this to find the components in the FLRW coordinates ($\chi.0,0,\hat{t}$) for the contravariant velocity vector $U^{\mu}$ of the point at $R$ .  
To get the time component, we divide eq \ref{1} by $d\hat{t} ^2$ with $d\omega = 0$ to obtain
\begin{equation}
	(\frac {ds}{d\hat{t}})^2 = 1 - a(t)^2 (\frac{d\chi}{d\hat{t}})^2 = (1 - \hat{V}^2) \equiv \frac{1}{{\gamma}^2}
\lbl{12'}
\end{equation}
 
To get the spatial component, we use the chain rule applied to eqs \ref{12} and \ref{12'}:
\begin{equation}
	\frac{d\chi}{ds} = \frac{d\chi}{d\hat{t}} \frac{d\hat{t}}{ds} =\ \frac{\hat{V}}{a}{\gamma}.
\end{equation}
Thus, the FLRW components of $U^{\mu}$ are
\begin{equation}
	U^{\mu} = (\frac{{\gamma} \hat{V}}{a},0,0,{\gamma}).
\lbl{u2}\end{equation}
with the time component listed last.  

The transformed coordinates are $(R,\theta,\phi,T)$.  To get $U^{\mu}$ in these coordinates, we make the spatial components be zero as required for it to be a vector velocity of the point $R$ (see the definitions under assumption iii).
This assumption means that a test particle attached to the radial coordinate will feel a force caused by the gravitational field, but will be constrained not to move relative to the coordinate.  Alternatively, a co-located free particle at rest relative to the radial point will be accelerated by the same force, but will thereafter not stay co-located.  
This means that the radial point is not inertial, except close to the origin.  

The time component of the transformed vector is $dT/ds = 1/cA$.  This makes the vector 
$U^{\mu}$ in the transformed coordinates be
\begin{equation}
	U^{\mu}= (0,0,0,\frac{1}{cA}).
\lbl{u1}\end{equation}
 
  Since $U^{\mu}$ is to be contravariant, its components will transform the same as $dT,dR$ in eq \ref{7}:

\begin{equation}
\begin{array}{rcl}
\frac{1}{cA}& = & \frac{1}{c}T_t{\gamma} + \frac{1}{a}T_\chi {\gamma} \hat{V},\\[.15cm]
0 & = & \frac{1}{c}R_t {\gamma} + \frac{1}{a} R_\chi {\gamma} \hat{V}. \end{array} \lbl{11}.
\end{equation}

Manipulating the second line of eq \ref{11} gives
\begin{equation}
\hat{V}  =  -\frac{a R_t}{cR_\chi}. \lbl{m13}\end{equation}

If we invert eq \ref{7}, we get
\begin{equation}
\begin{array}{l}
d\hat{t} = \frac{1}{D}(R_\chi dT - T_\chi dR),\\[2mm]
d\chi = \frac{1}{D}(- \frac{1}{c}R_t dT + \frac{1}{c}T_t  dR),
\end{array}  \lbl{iv}
\end{equation}

where 
\begin{equation}
 	D = \frac{1}{c}T_tR_\chi - \frac{1}{c}R_tT_\chi = \frac{1}{c}T_tR_\chi (1 + \hat{V}\frac{cT_\chi}{aT_t}),
\lbl{iv2}
\end{equation}
using eq \ref{m13}.

We can enter $d\hat{t}$ and $d\chi$ of eq \ref{iv} into the FLRW metric (eq \ref{1} ).  One way to make the line element $ds$ to be the same as FLRW is to equate these coefficients individually with those of eq \ref{5'}: 
\begin{equation}
	A^2 = \frac{1}{T_t^2} \left[ \frac {1-\hat{V}^2}{(1+\hat{V}\frac{cT_\chi}{aT_t})^2}\right],\lbl{in1}
\end{equation}
\begin{equation}
	B^2 = \frac{a^2}{R_\chi ^2} \left[ \frac{1-(\frac{cT_\chi}{aT_t})^2}{(1+\hat{V}\frac{cT_\chi}{aT_t})^2}\right],\lbl{in2}
\end{equation}
and
\begin{equation}
	C = -\frac{a}{T_tR_\chi } \left[ \frac{\hat{V}+\frac{cT_\chi}{aT_t}}{(1+\hat{V}\frac{cT_\chi}{aT_t})^2} \right].\lbl{in3}
\end{equation}
If we put $ds=0$ in eq \ref{5'}, we obtain a coordinate velocity of light $v_p$:
\begin{equation}
	\frac {v_p}{c} =  \left[\frac{\partial R}{c\partial T}\right]_s = -\frac{C}{B^2} \pm \sqrt{(\frac{C}{B^2})^2 +  \frac{A^2}{B^2}}
\lbl{2'}
\end{equation}

The equations for $A$, $B$, and $v_p$ simplify for a diagonal metric ($C = 0$). From eq \ref{in3} we get
\begin{equation}
\frac{cT_\chi}{aT_t} = -\hat{V}
\lbl{diag}\end{equation} 
So rigidity gives a relation of $dR$ to $V$ (eq \ref{m13}), and diagonality gives a relation of $dT$ to $V$ (eq\ref{diag}).
 Thence,  eqs \ref{in1}, \ref{in2}, and \ref{2'} become
\begin{equation} 
	A = \frac{{\gamma}}{T_t} = \frac{t_T}{{\gamma}}
\lbl{in1'}\end{equation}
\begin{equation}
	B =  \frac{a{\gamma}}{R_\chi} = \frac{a\chi_R}{{\gamma}}
\lbl{in2'}\end{equation}
\begin{equation}
	\frac {v_p}{c} = \frac{A}{B},
\lbl{in2"}\end{equation}
where we have used eq \ref{iv} with $C = 0$ to obtain the inverse partials.
This metric becomes the physical Minkowski ($\hat{M}$) when $A \rightarrow 1, B \rightarrow 1, C \rightarrow 0 $ and $ar \rightarrow R$.  
Thus, the coordinate light speed for the stationary coordinates starts at $c$ for
 $R = 0$ where $T_t = {\gamma} = R_\chi/a = 1$, and increases by the ratio $A/B$ as $R$ increases  .

We can say something about the physicality of the coordinates with a generalization of criteria ($iv$) developed by Bernal et al\cite{B}.
They developed a theory of fundamental units based on the postulate that two observers will be using the same units of measure when each measures the other's differential units at the same space-time point compared to their own and finds these reciprocal units to be equal.  We generalize this by stating that if one coordinate of a system represents the reading on a physical instrument, so must the corresponding coordinate of the other  reciprocal system represent readings on the same type of physical instrument with the same units.  Thus, 
even if $A \ne 1$, $dR$ will be physical if $C = 0$ and $B = 1$ because then $R_\chi /a = a \chi_R = {\gamma}$ (eq \ref{in2'}) and $dR$ is physical because it uses the same measure of distance as $ad\chi$, which we assume is physical. 
Then $R$ uses physical units when $dR$ is integrated indefinitely out to the visible horizon.

At this point we would like to examine quantitatively how far from the $\hat{M}$ metric our stationary  metric is allowed to be in order for its coordinates to reasonably represent physical measurements.  We can consider the coefficients $A$, $B$, and $C$ one at a time departing from their value in the $\hat{M}$ metric. Thus, let us consider the physical distance case $B=1,C=0$ and examine the possible departure of the time rate in the transform from that physically measured. Then, from eqs \ref{in1'}: $T_t = {\gamma}/A, t_T = {\gamma} A$.  
Thus, $1-A$ represents a fractional increase from ${\gamma}$ in the transformed time rate $T_t$, and thus the fractional increase from the physical $T_t$ of an inertial rod at rest with the stationary coordinates at that point.  
We can make a contour of constant $A$ on our world map to give a limit for a desired physicality of the stationary time.

\section{Stationary physical distance using a diagonal metric} \lbl{V}

For diagonal coordinates with physical $dR$ for all $t$ and $\chi$, $B=1$, so eq \ref{in2} becomes
\begin{equation}
R_\chi = a{\gamma}
\lbl{r1}
\end{equation}
By integration we find
\begin{equation}
	R = a\int_0^\chi {{\gamma} \partial\chi_t},
\lbl{r2}
\end{equation}
where the subscript on $\partial\chi_t$ represents integration of $\chi$ at constant $t$.   Partial differentiation with respect to $t$ gives
\begin{equation}
	R_t = c\dot{a} \int_0^\chi{{\gamma}\partial\chi_t} +a\int_0^\chi{{\gamma}_t\partial\chi_t},
\lbl{r3}
\end{equation}
where the dot represents differentiation by $cdt$.  We can then find $\hat{V}$ from eqs \ref{m13}, \ref{r1}, and \ref{r3} as
\begin{equation}
	\hat{V} = -\frac{R_t}{c{\gamma}} = -\frac{1}{c{\gamma}}\left[ c\dot{a} \int_0^\chi{{\gamma}\partial\chi_t} +a\int_0^\chi{{\gamma}_t\partial\chi_t}\right].
\lbl{r4}
\end{equation}
This is an integral equation for $\hat{V}$.  It can be converted into a partial differential equation by multiplying both sides by ${\gamma}$ and partial differentiating by $\chi$:
\begin{equation}
	{\gamma}^2 \left[ \hat{V}_x + \frac{a}{c} \hat{V} \hat{V}_t \right] = -\dot{a} = -\frac{1}{ c} \frac {da}{dt}.
\lbl{r5}
\end{equation}
General solutions of eq \ref{r5} for $\hat{V}(t,\chi)$ are found in the Appendix. $T(t,\chi)$, and $R(t,\chi)$ are also found there by their dependence on $\hat{V}(t,\chi)$.  This will complete our search for a transform from FLRW that will satisfy the General Relativity Field Equation with a physical $R$ to all radial events.

\section{Interpretation}\lbl{I}

We can use these solutions to show on a $R,T$ plot the paths of galaxies ($\chi =$ constant) and photon ($ds = 0$).  We will use $\Omega$ as defined by Peebles\cite{P} (App \ref{sol}).  Thus $\Omega = 1$ is a flat universe ($k = 0$) with no cosmological constant ($\Lambda = 0)$.  These are shown in Figs. 1 and 2.  An approximate upper limit of physicality is shown by the dashed line for $A = 1.05$.  Below this line, $T$ is close to physical. 
$R$ vs $T$ at $t=0$ provides a horizon, where the visible universe has a finite physical distance, but where $T$ is non-physical. \\

[Insert Fig 1]\\

Fig 1 coordinates are both physical, the physical distance $R$ to a galactic point at $\chi$ characterized by the red shift $z = -1 + (t_0/t)^{2/3}$ and the time at the origin (or on any galactic point) $t/t_0$.  The light we see now at the origin originates from a galactic point when its path crosses the light path, where the physical distance is as shown.  Notice that the light comes monotonically towards us even from the farthest galactic point, but its coordinate speed slows down the farther it is away in these coordinates where the clocks measuring $t$ are on moving galactic points whereas the rulers measuring $R$ are on the stationary frame. Thus,  $(a\partial \chi/\partial t)_s = c$ when clocks and rulers are on the same frame, but $(\partial R/\partial t)_s$ is a coordinate light speed not equal to $c$ (except close to the origin where t approaches T) because $R$ and $t$  are on different frames  This is
 much like the coordinate light speed of the Schwartszchild coordinates\cite{We1} where clocks and distances are at different locations.\\

[Insert Fig 2]\\

Although the distance to the origin in Fig 2 is physical, the co-located time coordinate $T$ is not, but comes close to it near the space origin.  For distances and times below the physicality line $A = 1.05$, $T_t$ is less than the physical time by only $5 \%$ on a co-located stationary inertial frame. 

As such, the physical interpretation seems very clear.  A galactic point seems to start at $R = 0,T = 0$ and moves out to the time it emits its light that can be later seen at the origin at $T = t_0$ . The farthest galactic points travel out the fastest and release their light the earliest, but even for the most remote galactic point, the distance never gets greater than $.58 c_0t_0$, which can be considered to be the size of the visible universe when the farthest light was emitted.  

Near the horizon, some oddities occur in this interpretation; when $t \rightarrow 0$, $T$ comes close but is not zero for most galactic paths, but also is not physical at that point.  $R$ uses physical rulers at that point, but also is not quite zero. This discrepancy becomes the largest for the farthest galactic points. 
The lack of physicality near the horizon is presumably what causes the transform to show 
non-zero times and distances at $t = 0, {\gamma} = \infty$ and the farthest galactic points to go close to
the stationary coordinate light speed which approaches $1.3c$ near the horizon.  
Of course, if there were to be a different stationary transform with constant inertial light speed that had a rigid physical distance, it might have a different universe radius, but this seems unlikely. (A possible alternative with different assumptions is explored in \cite{RCF}).\\

[Insert Fig 3]\\

The above interpretation is strengthened by examining the physical coordinates of an expanding empty FLRW  universe, a solution that was first published by Robertson with $a = ct \ $\cite{R2}:
\begin{equation}
\begin{array}{l}    R = ct\sinh{\chi} ,\\ T = t \cosh{\chi}. \end{array} \lbl{51} \end{equation} 
Robertson showed that these transformed coordinates obeyed the Minkowski metric for all $t,\chi$, as is required of physical coordinates in empty space. This solution is plotted in Fig 3 with the same format as Fig 2.  The paths are very similar except that all the paths are straight with no universe density to curve them, and the galactic paths do not have a gap at the singularity at $t = 0$. Notice that along the light path that $\chi_p = \ln{(t_0/t )}$, and $R_p = ct_0(1 - t^2/t_0^2)/2$ so that at $t = 0$, the extent of the visible universe is $R_p = ct_0/2$.  In terms of the clocks at the origin that read $t$, light seems to come monotonically towards us from this distance, the coordinate light speed $(\partial R/\partial t)_s$ slowing to zero at $t = 0$. 

It's natural to wonder what is outside the horizon.  In this empty universe, all galactic points are within the horizon. Outside the horizon not only are there no galactic points, there is $no \ space$  when viewed from a stationary frame.  Like a Fitzgerald contraction all differential radial distances shrink to zero at the horizon as the galactic points approach the light speed.  From one of those points, it would have a different stationary frame and so would see a finite space in its vicinity and would see a different horizon.  
  
For a finite universe energy density, many higher values of $\chi$ are not included inside the horizon.  All of this is clouded by the nonphysicality of $T$ close to the horizon.
But it would appear that all galactic matter starts outward from $R = 0$ at $T = 0$ traveling close to the light speed near the horizon, the largest $\chi$ traveling the fastest.  As each $\chi_1$ is slowed down by the inward gravitational pull of the mass of galaxies inside it, it departs from the horizon after a time $T_1(\chi_1)$ at a distance $R_1(\chi_1)$, where it can emit light that will be the earliest light visible at the origin at time $T = t_1$, so $\chi_1 = 3 (t_1/t_0)^{1/3}$ for $\Omega = 1$ (see eq \ref{r13}).  Those with $\chi > \chi_1$ will not be visible at the origin at $T = t_1$, because they must be inside the horizon to emit light, presumably because of the singularities in ${\gamma}$ and in energy density along the horizon.  
 Thus, only a limited number of galactic points can be visible at the origin at the present time, those with 
$t_1 \leq t_0$.

This is a different picture from a universe that is full of galactic points beyond a visible horizon, but are invisible only because
there is not enough time to see them.  I'm suggesting that like the empty universe, there are no galactic points and no space beyond the horizon when viewed from a stationary frame.

\section{Newtonian gravity for flat space ($\Omega = 1$) close to the origin}

In Appendix \ref{s2.1.1}, the acceleration vector $A^\mu$ of a point on $R$ is found in FLRW coordinates and in stationary coordinates. The latter is solved for a flat universe ($k = 0$) with no dark energy $\Omega = 1$ in normalized coordinates 
($\chi a_0/ct_0 \rightarrow x, t/t_0 \rightarrow t, 
$ see \ref{gflat}).  For small $u = x /t^{1/3}$, the gravitational acceleration $g$ goes to zero as $-2u/9t = -2R/9 $.  
Since small $u$ is the region with physical coordinates, it is interesting to express $g$ and $R$ in unnormalized coordinates:
\begin{equation}
	-g \rightarrow  \frac{2}{9} \frac{c}{t_0} \frac{R}{c t_0} =   \frac{4\pi \rho_0 G R}{3} =  \frac{GM_0}{R^2},
\lbl{gB1}\end{equation}
where we have used $\rho = \rho _0 (a_0/a)^3$ in eq \ref{g13} and $M_0$ as the present universe mass inside the radius $R$.  Thus, the gravitational force in the stationary coordinates used here near the origin is the same as Newtonian gravity,  
 as one might expect if both $T$ and $R$ are physical near the origin and if $V << c$.

\section{Conclusion}

We have found a transform to stationary coordinates $T,R$ from the FLRW coordinates $t,\chi$ for radial events that has a physical radial distance to each event.  Although the time of the transform is not physical for all events, it is physical close to the space origin.  When the physical radius is plotted against the physical time at the origin, the coordinate light speed $dR/dt$ slows down the farther out it is in the universe, much like the coordinate light speed of Schwartszchild coordinates as it approaches the event horizon.  But, since the time and distance of the stationary coordinates are both local on the same frame, the light speed in these coordinates remains fairly constant out to the far reaches of the universe.  Near the origin in these coordinates, the gravitational acceleration becomes Newtonian for a flat universe ($\Omega = 1$).  
The stationary coordinates indicate that galactic points originate at $T = 0$, move out to a finite distance in a finite time when it releases light that we can see at the present time.  For the most distant galactic points the distance at emission is about half of the so-called time-of-flight distance $c(t_0-t_e)$.
An expanding horizon occurs in the stationary coordinates where the radial velocity of the galactic points approaches the light speed, outside of which there are no galactic points, and no space.  At the time of the release of the earliest light that we could see today, the horizon was at $R = .58 ct_0$ in the stationary coordinates for a flat universe ($\Omega = 1$) with no dark energy.

\section*{Appendix}

\appendix

\addcontentsline{toc}{section}{Appendix}

\section{Stationary coordinates for any $a(t)$}
\subsection{Solution for $\hat{V}$}
Eq \ref{r5} can be solved as a standard initial-value problem.  Let $W \equiv -\hat{V}$.  Eq \ref{r5} becomes
\begin{equation}
W_\chi - \frac{a}{c}WW_t = \frac{1}{c} \frac {da}{dt}(1-W^2)
\lbl{g5}\end{equation}
Define a characteristic for $W(t,\chi)$ by
\begin{equation}
(\frac{\partial t}{\partial \chi})_c = -\frac{a}{c}W
\lbl{g6}\end{equation}
so
\begin{equation}
(\frac{\partial W}{\partial \chi})_c = \frac{1}{c} \frac {da}{dt}(1-W^2)
\lbl{g7}\end{equation}
(The subscript $c$ here indicates differentiation along the characteristic).  If we divide eq \ref{g7} by eq \ref{g6} we get
\begin{equation}
(\frac{\partial W}{\partial t})_c = -\frac{1}{a} \frac {da}{dt}\frac{(1-W^2)}{W}
\lbl{g8}\end{equation}
This can be rearranged to give
\begin{equation}
\frac{W (\partial W) _c}{W^2-1} = \frac{(\partial a) _c}{a}
\lbl{g9}\end{equation}
This can be integrated along the characteristic with the boundary condition at $\chi=0$ that $W = 0$ and $a = a_c$:
\begin{equation}
1-W^2 = \frac{a^2}{a^2_c} = \frac{1}{{\gamma} ^2}.
\lbl{g10}\end{equation}
This value for $W$ (assumed positive for expanding universe) can be inserted into eq \ref{g6} to give
\begin{equation}
(\frac{\partial t}{\partial \chi})_c = -\frac{a}{c}\sqrt{1-\frac{a^2}{a^2_c}}
\lbl{g11}\end{equation}
We can convert this to a differential equation for $a$ by noting that $c(\partial t)_c = (\partial \hat{t})_c = \frac{1}{\dot{a}} (\partial a) _c$
\begin{equation}
(\frac{\partial a}{\partial \chi})_c = -a \dot{a}\sqrt{1-\frac{a^2}{a^2_c}}.
\lbl{g12}\end{equation}
$\dot{a}$ can be found as a function of $a$ from the well known solution of the GR field equation for the FLRW universe\cite{P}:
\begin{equation} 
\dot{a} = a \sqrt{\frac{8\pi G}{3c^2}\rho -\frac{k}{a^2} + \Lambda /3 }
\lbl{g13} \end{equation} 
 where $\rho$ is the rest mass density of the ideal liquid assumed for the universe that can be found as a function of $a$ (see App \ref{A4});  G is the gravitation constant; 
 $k$ is the constant in the FLRW metric (eq \ref{2"}); and $\Lambda$ is the cosmological constant possibly representing the dark energy of the universe.

Eq \ref{g12} can be integrated along the characteristic with constant $\alpha_c$, starting with $\alpha = \alpha_c$ at $\chi= 0$.  This will give $\chi = X(\alpha,\alpha_c)$.  This can be inverted to obtain $\alpha_c(\alpha,\chi) $.  When this is inserted into eq \ref{g10}, we have a solution to eq \ref{g5} for $W(\alpha,\chi)$.  Integration of eq \ref{g13} gives $a$  and thus $W$ as a function of time $t,\chi$. 

\subsection{Obtaining $T,R$ from $\hat{V}$} \lbl{pd}

Eqs \ref{12}, \ref{m13}, and \ref{diag} show that
\begin{equation}
W = -\frac{a}{c}(\frac{\partial \chi}{\partial t})_R = \frac{a}{c}\frac{R_t}{R_\chi} = \frac{c}{a}\frac{T_\chi}{T_t}
\lbl{g16'}\end{equation}
so
\begin{equation} 
T_\chi - \frac {a}{c}WT_t = 0.
\end{equation}
Thus $T$ has the same characteristic as $W$ (eq \ref{g6}), so that $(\partial T/\partial \chi)_c = 0$, and $T$ is constant along this 
characteristic:
\begin{equation}
T(t,\chi) = T(t_c,0) = t_c \equiv t(\alpha_c(t,\chi))
\lbl {g16}\end{equation}
where $t(\alpha)$ is given by the integration of eq \ref{g13} and $\alpha_c(\alpha(t),\chi)$ is found by inverting the integration of eq \ref{g13}.  This gives us the solution for $T(t,\chi)$ and $A$.  
\begin{equation} 
A = \frac{{\gamma}}{T_t} = \frac{a_c}{a}(\frac{\partial t}{\partial t_c})_\chi = \frac{a_c}{a}\frac{d a_c/d t_c}{d a/dt} (\frac{\partial a}{\partial a_c}) _\chi .
\end{equation}

The solution for $R$ can be obtained by integrating eq \ref{r2}, using ${\gamma}$ from eq \ref{g10} and $a_c(t,\chi)$ from eq \ref{g13}:
\begin{equation}
R(t,\chi) = a\int_{0}^{\chi}{{\gamma}  \partial \eta _t} = \int_{0}^{\chi}{a_c(t,\eta) \partial \eta _t} .
\lbl{g17}\end{equation}

Alternatively, for ease of numerical integration we would like to integrate $dR$ along the same characteristic as $T$ and $W$.  This can be obtained from the PDE
\begin{equation}                   
(\partial R/\partial \chi)_c = R_\chi + R_t (\partial t/\partial \chi)_c
\end{equation}
If we insert the values for these three quantities from eqs \ref{r1}, \ref{g16'}, and \ref{g6} , we get
\begin{equation}
(\partial R/\partial \chi)_c = {\gamma} a + (cW/a) {\gamma} a(-aW/c) = a/{\gamma} = a^2/a_c.
\lbl{g19}\end{equation}

\section{Similarity solutions for flat universe ($\Omega = 1$)}\lbl{fu}

But I have found a simpler integration of eq \ref{r5} for the special case of $\Omega = 1$ where the GR solution is $a = a_0(t/t_0)^{2/3}$\cite{P}.   
To simplify notation, let us normalize:  $t/t_0 \rightarrow t$, $a/a_0 = \alpha = t^{2/3}$, and $
\chi a_0/ct_0 \rightarrow x$, $T/t_0 \rightarrow T $, $R/ct_0 \rightarrow R$, and let $W = -\hat{V}$.  

\subsection{Ordinary differential equation for V or -W}

Eq \ref{r5} then becomes
\begin{equation}
	W_x - t^{2/3}WW_t = \frac{2}{3}t^{-1/3}(1-W^2)
\lbl{r6}
\end{equation}
This can be converted into an ordinary differential equation (ODE) by letting
\begin{equation}
	u \equiv x/t^{1/3}
\lbl{r7}
\end{equation}
so that eq \ref{r6} becomes
\begin{equation}
	W'(1+\frac{uW}{3}) = \frac{2}{3}(1-W^2),
\lbl{r8}
\end{equation}
where the prime denotes differentiation by $u$.

\subsection{Ordinary differential equation for T and R}
Similarly we can find ODE's for $T$ and $R$ by defining:
\begin{equation}
	T/t \equiv q(u),	
\lbl{r9}
\end{equation}
and
\begin{equation}
	R/t \equiv s(u),
\lbl{r10}
\end{equation}
where $q(u)$ and $s(u)$, from eqs \ref{diag} and \ref{m13}, are given by the coupled ODE's:
\begin{equation}
	q'(1+\frac{uW}{3}) = qW,
\lbl{r11}
\end{equation}
and
\begin{equation}
	s'(W+\frac{u}{3}) = s
\lbl{r12}
\end{equation}
It is useful to find that $q = {\gamma}^{3/2}$, $s = {\gamma} (u+3W)/3$, $s' = {\gamma}$, and $A = {\gamma}/T_t = (1+uW/3)/{\gamma} = v_p/c $; so $T = t{\gamma}^{3/2}$, and $R = t{\gamma} (u+3W)/3$.

For small values of $u$, $W = 2u/3$, $q = 1 + u^2/3$, $s = u$, $v_p/c = 1+\bigcirc{(W^4)}$, and $R = t^{2/3}x = ax$.  The light speed $v_p$ measured on $T,R$ remains close to $c$ out to large $R$.  
We also note that $T_t \rightarrow 1 + W^2/4$, which is slower than a Lorentz requirement of $T_t \rightarrow {\gamma} \rightarrow 1 + W^2/2$. 

As $t \rightarrow 0$, $u \rightarrow \infty$, ${\gamma} \rightarrow \kappa u^2$, $W \rightarrow 1 - 1/(2\kappa^2 u^4) $, $q \rightarrow \kappa^{3/2}u^3 $, and $s \rightarrow  \kappa u^3 /3$.  $T$ and $R$ both remain finite at this limit with $T \rightarrow \kappa^{3/2} x^3 $, and $R \rightarrow  \kappa x^3 /3$ with $T/R \rightarrow 3\kappa^{1/2}$.  $\kappa$ is difficult to determine from the numerical integration because of the singularity at large $u$, but my integrater gives $\kappa = .0646$.     
The fact that $T$ does not go to zero when $t$ goes to zero results from equating $T$ with $t$ at $t = 1$ and not at $t = 0$.  

The distance $R$ and time $T$ can be found from the numerical integration of the coupled ODE's. 
The paths of galactic points are those for constant $x$.  The path photons have taken reaching the origin at $t_1$ is found by calculating $x_p$ vs $t$ and using the transform to $T,R$.  Thus, for $\Omega = 1$
\begin{equation}
	x_p = \int_{t_1}^t{\frac{c}{a} dt} = 3(t_1^{1/3}-t^{1/3} )
\lbl{r13}
\end{equation}
For light arriving now, $t_1 =1$, the value of $u_p$ becomes
\begin{equation}
	u_p = 3(\frac{1}{t^{1/3}} -1) .
\lbl{r13'} \end{equation}
Note that $x_p \rightarrow 3$ as $t \rightarrow 0$ at the beginning of the path for photons. This makes the horizon at the earliest time of release of light visible today be $.581 ct_0$.

\section{Gravitational field } \lbl{s2.1.1}

\subsection{Gravitational field in the FLRW and stationary coordinates} 

We wish to find the components of the radial acceleration of a test particle located at R in the stationary  system.  We will do this by calculating the FLRW components of the acceleration vector and find the transformed components by using the known diagonal transforms.  For the FLRW components, we will use the metric
\begin{equation}
	ds^2 = d\hat{t}^2 - a^2 d\chi^2 -a^2r^2 d\theta ^2 -a^2r^2\sin^2{\theta} d\phi^2 .
\end{equation}
Let
\begin{equation}
	x^1 = \chi,~ x^2 = \theta,~ x^3~ = \phi,~x^4 = \hat{t} = ct, \lbl{m1}\end{equation}
and the corresponding metric coefficients become
\begin{equation}
g_{44} = 1,~g_{11} = -a^2,~g_{22} = -a^2r^2,~g_{33} =
-a^2r^2\sin^2{\theta}. \lbl{m2}\end{equation}

For any metric, the acceleration vector for a test particle is 
\begin{equation}
	A^{\lambda} = \frac{dU^{\lambda}}{ds} + \Gamma_{\mu \nu}^{\lambda} U^{\mu} U^{\nu}, 
\lbl{m3}\end{equation}
where the $\Gamma$'s are the affine connections and  $U^\lambda$ is the velocity vector of the test particle.  In our case the test particle is at the point $R$ on the transformed coordinate,  but not attached to the frame so that it can acquire an acceleration. Instantaneously, it will have the same velocity as the point on the transformed coordinate, and its velocity and acceleration vectors will therefore transform the same as the point (eq \ref{11}).

We will be considering accelerations only in the radial direction so that we need find
affine connections only for indices 1,4.  The only non-zero
partial derivative with these indices is
\begin{equation}
\frac{\partial g_{11}}{\partial x^4} = -2a\dot{a}. \lbl{m5}\end{equation}
 
The general expression for an affine connection for a diagonal metric is
\begin{equation}
 \Gamma^\lambda_{\mu\nu}   = \frac{1}{2g_{\lambda\lambda}}\left[ \frac{\partial g_{\lambda\mu}}{\partial
 x^\nu} + \frac{\partial g_{\lambda\nu}}{\partial
 x^\mu} - \frac{\partial g_{\mu\nu}}{\partial
 x^\lambda}\right]. \lbl{m6}\end{equation}
The only three non-zero affine connections with 1,4 indices
are
\begin{equation}
\Gamma^4_{11} = a\dot{a},~\Gamma^1_{41} = \Gamma^1_{14} =
\frac{\dot{a}}{a}. \lbl{m7}\end{equation}
The acceleration vector in FLRW 
coordinates of our test particle moving at the same velocity as a point on the transformed frame becomes 
\begin{equation}
\begin{array}{l}
A^{\hat{t}} = \frac{dU^4}{ds} + {\Gamma}^4_{11}U^1 U^1, \\[2mm]
A^\chi = \frac{dU^1}{ds} + \Gamma^1_{41}(U^4 U^1 + U^1 U^4).  \end{array} \lbl{m8}\end{equation}
Using $U^4$ and $U^1$ in eq \ref{u2}, we find
\begin{equation}
\begin{array}{l}
A^{\hat{t}} = {\gamma}\left( \frac{\partial {\gamma}}{\partial \hat{t}}\right)_{\! R}
 + a\dot{a}\frac{{\gamma}^2\hat{V}^2}{a^2} = {\gamma}^4\hat{V}\left( \frac{\partial \hat{V}}{\partial \hat{t}}\right)_{\! R}
 + \frac{\dot{a}}{a}{\gamma}^2\hat{V}^2,\\[2mm]
A^\chi  = {\gamma}
\left( \frac{\partial}{\partial \hat{t}}({\gamma} \hat{V}/a) \right)_{\! R}
+ 2\frac{\dot{a}}{a}\frac{{\gamma}^2\hat{V}}{a} = \frac{{\gamma}^4}{a}   \left( \frac{\partial \hat{V}}{\partial \hat{t}}\right)_{\! R} + 
\frac{\dot{a}}{a^2}{\gamma}^2\hat{V}. \end{array} \lbl{m9}\end{equation}
Since the acceleration vector of the test particle at $R$ in the stationary coordinates will be orthogonal to the velocity vector, it becomes
\begin{equation}
\begin{array}{l}
A^T = 0,\\[2mm]
A^R = \frac{DU^R}{Ds}  \equiv -g/c^2. \end{array} \lbl{m23}\end{equation}
$A^R$ is the acceleration of a point on the $R$ axis, so the gravitational field affecting objects like the galactic points is the negative of this.  
$g$ is defined so that $mg$ is the force acting on an object whose mass is $m$. 
Close to the origin, $g = \frac{d^2R}{dT^2}$, the normal acceleration.  Since the vector $A^\lambda$ will transform like $dT,dR$ (eq \ref{7}):
\begin{equation}
	A^R = \frac{1}{c}R_tA^{\hat{t}} + R_\chi A^\chi 
\end{equation}
so that
\begin{equation}
	-g/c^2 = \left[ {\gamma}^4 \hat{V}\left( \frac{\partial \hat{V}}{\partial \hat{t}}\right)_{\! R} +
\frac{\dot{a}}{a}{\gamma}^2\hat{V}^2 \right] \frac{1}{c}R_t + \left[  \frac{{\gamma}^4}{a}\left( \frac{\partial \hat{V}}{\partial \hat{t}}\right)_{\! R} +
\frac{\dot{a}}{a^2}{\gamma}^2\hat{V} \right] R_\chi.
\end{equation}

With the use of eq \ref{m13}, this can be simplified to
\begin{equation}
	-g/c^2 = \frac{R_\chi}{a} \left[ {\gamma}^2 \left( \frac{\partial \hat{V}}{\partial \hat{t}}\right)_{\! R} + \frac{\dot{a}}{a} \hat{V} \right] .
\lbl{g} \end{equation}
The acceleration $g$ can be thought of as the gravitational field caused by the mass of the surrounding galactic points, which balances to zero at the origin, where the frame is inertial, but goes to infinity at the horizon.  It is the field which is slowing down the galactic points (for $\Lambda = 0$).

\subsection{Gravitational force for flat universe ($\Omega = 1$)}\lbl{gflat}

If we insert the values of $V$, $R$, and $a/c$ in Appendix \ref{s2.1.1}, eq \ref{g}, we obtain
\begin{equation}
g = -\frac{s'}{t}\left[ {\gamma}^2 W'(\frac{u}{3}+W)-\frac{2W}{3} \right]
\lbl{fug}\end{equation}
where $g$ has the units $c_0/t_0$.  The insertion of $W'$ and $s'$ into eq \ref{fug} gives
\begin{equation}
	-g = \frac{2}{9 {\gamma} t}\left[ \frac{u}{1+uW/3} \right],
\lbl{gB} \end{equation}

\section{Solution of the FLRW universe field equation} \lbl{sol} \lbl{A4}
Peebles has shown a convenient was to represent eq \ref{g13}\cite[p312]{P}, the solution of the field equation for the FLRW metric in a universe filled with an ideal liquid.  He defines
\begin{equation}
	\Omega \equiv \rho _0  \frac{8\pi G }{3c^2H_0^2}, 
\lbl{a3}
\end{equation}
and
\begin{equation}
	\Omega _r \equiv \frac{-k}{H_0 ^2a_{0}^2},
\lbl{a4}
\end{equation}
and
\begin{equation}
	\Omega _{\Lambda} \equiv \frac {\Lambda}{3 H_0 ^2}.
\lbl{a5}
\end{equation}

$\rho_0$, $H_0$, and $a_0$ are the energy density, Hubble constant, and universe scale factor, resp., at the present time $t_0$.  
For very small $a$ there will also be radiation energy term $\Omega_R \approx 2 \cdot 10^{-5}$\cite{P}.   

Let $\alpha = a/a_0$.
 It is found by the solution of the ordinary differential equation:
\begin{equation} 
 \frac {1}{\alpha}\frac{d\alpha}{d\hat{t}} \equiv H = H_0 E(\alpha), 
\end{equation}
where the normalized Hubble ratio $E$ is
\begin{equation} 
	 E(\alpha) = \sqrt{\frac{\Omega_R}{\alpha^4} + \frac{\Omega}{\alpha ^3} + \frac {\Omega _r}{\alpha ^2} + \Omega _{\Lambda}}.
\lbl{a7}
\end{equation}
The $ \Omega $s are defined so that
\begin{equation}
	\Omega _R + \Omega + \Omega_r + \Omega _{\Lambda} = 1.
\end{equation} 
At $t = t_0$: $\alpha = 1$ and $E = 1$.

The cosmic time $t$ measured from the beginning of the FLRW universe becomes
\begin{equation}
	cH_0 t =  \int_0^{\alpha}\frac {d\alpha}{\alpha E} . 
\lbl{a9}
\end{equation}

For a flat universe with $\Omega = 1$ and $\Omega_R =\Omega_r = \Omega_{\Lambda} = 0$: 
\begin{equation} 
\alpha = (t/t_0)^{2/3}, \\\ t_0 = \frac {2}{3cH_0}.
\end{equation}

\section*{Acknowledgments} \lbl{ack}
\addcontentsline{toc}{section}{Acknowledgments}

I wish to acknowledge the invaluable help given by Paul Fife, University of Utah Mathematics Dept.

\section*{Figures} 
\addcontentsline{toc}{section}{Figures}

\begin{figure}[tbp] 
  \centering
  \includegraphics[bb=51 108 742 504,width=6.67in,height=4.25in,keepaspectratio]{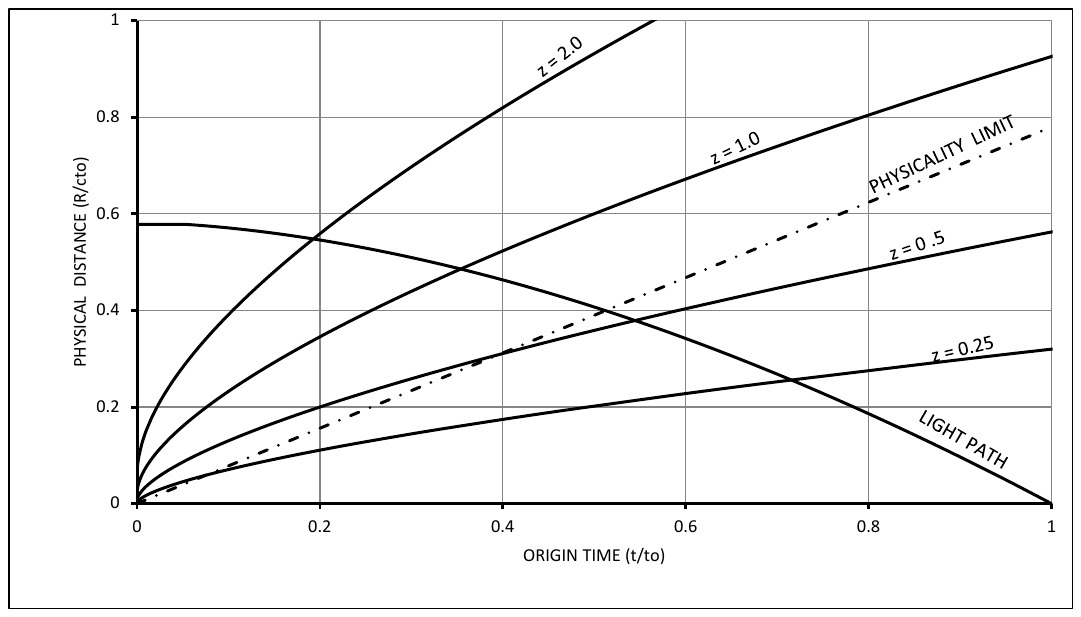}
  \caption{Physical distance $R/c_0t_0$ for a flat universe ($\Omega = 1$) vs physical time at the origin $t/t_0$ (or galactic time).  The paths of galactic points representing galaxies are characterized by their red shift $z$. The transformed time (not shown) is close to physical only below the PHYSICALITY LIMIT line where $A = 1.05$.  Light comes towards the origin along the LIGHT PATH from all radial points of the universe, traveling slower than its present speed in these coordinates where the clocks are on different frames from the rulers (like the Schwarzschild coordinates).  The galactic paths show the expanding universe in physical coordinates, some traveling faster than the light speed in these  coordinates.}
  \label{fig:RT10F1}
\end{figure}

\begin{figure}[tbp] 
  \centering
  \includegraphics[bb=50 226 562 566,width=5.67in,height=3.76in,keepaspectratio]{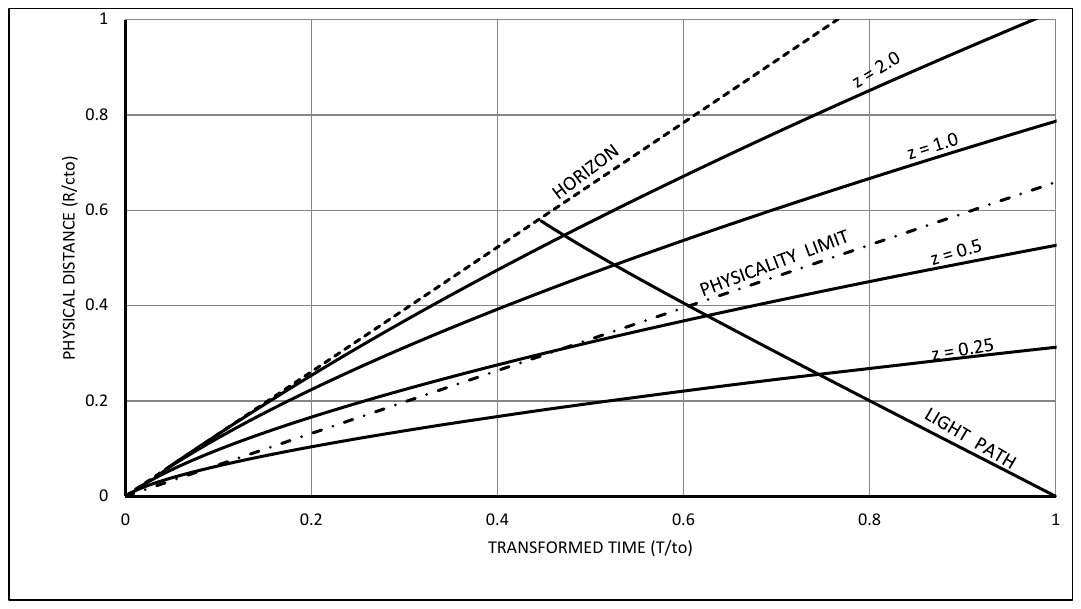}
  \caption{Physical distance ($R/c_0t_0$) for a flat universe $\Omega = 1$) vs the stationary local time $T/t_0$, which is close to physical below the PHYSICALITY LIMIT line ($A = 1.05$). 
A galactic point starts close to the physical origin at $T = 0$ and travels out to where it may emit light when it crosses the LIGHT PATH that will travel close to the physical light speed and be visible later at the physical origin at $T = t_0$.  The line labeled HORIZON represents singularities where $t = 0$ and ${\gamma} = \infty$, beyond which there are no galactic points, and no space.}
  \label{fig:RT10F2}
\end{figure}

\begin{figure}[tbp] 
  \centering
  \includegraphics[bb=101 268 508 537,width=5.67in,height=3.74in,keepaspectratio]{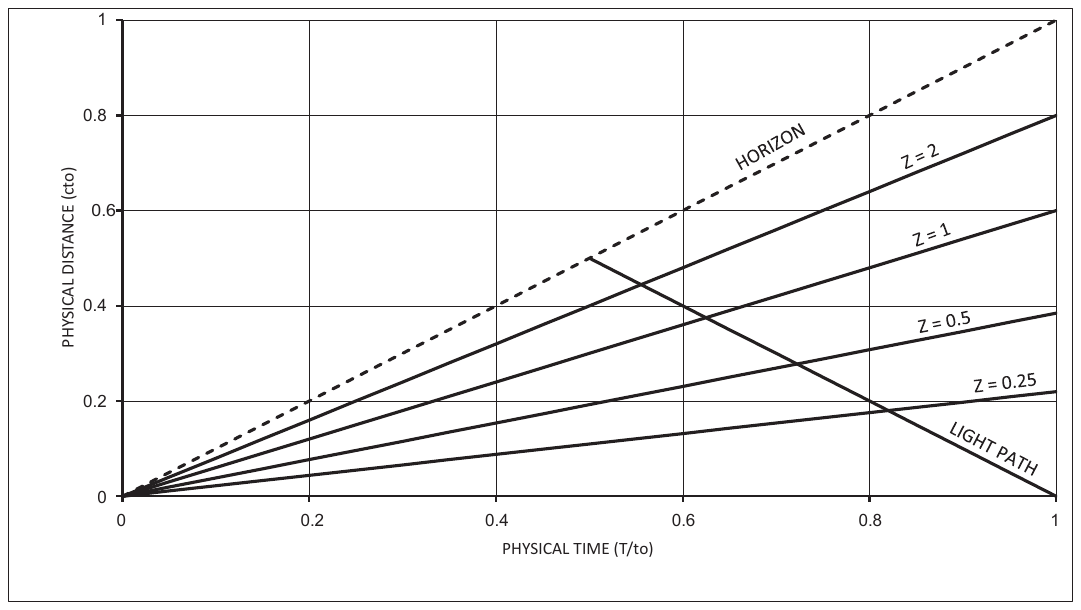}
  \caption{Physical distance($R/ct_0$) for the empty expanding universe ($\Omega = 0, \Omega _r = 1) $  plotted against the transformed time ($T/t_0$) on clocks attached at $R$ for various galaxy paths (labeled by their red shift $z$) and for the light path that arrives at the origin at $T = t_0$.  The horizon is the locus of points where $t = 0$.  All lines are straight and physical, since there is no space curvature.  The remotest galactic point travels from the origin at $T = 0$ out to $ct_0/2 $ at the light speed $c$.  There is no space outside the horizon to contain any galactic points.}
  \label{fig:RT01F3}
\end{figure}

\end{document}